# A New Compression Based Index Structure for Efficient Information Retrieval

**Md. Abdullah al Mamun[1], Md. Hanif[2], Md. Rakib Uddin[3], Tanvir Ahmed[4], Md. Mofizul Islam[5]**

[1,2,5]Department of ICT, Mawlana Bhashani Science and Technology University (MBSTU).
Bangladesh.
[4]Department of Computer Science, American International University-Bangladesh (AIUB).
Kemal Ataturk Avenue, Banani, Dhaka, Bangladesh.
[3]Department of Pharmaceutical Chemistry, University of Dhaka.
Dhaka, Bangladesh

## ABSTRACT

Finding desired information from large data set is a difficult problem. Information retrieval is concerned with the structure, analysis, organization, storage, searching, and retrieval of information. Index is the main constituent of an IR system. Now a day exponential growth of information makes the index structure large enough affecting the IR system's quality. So compressing the Index structure is our main contribution in this paper. We compressed the document number in inverted file entries using a new coding technique based on run-length encoding. Our coding mechanism uses a specified code which acts over run-length coding. We experimented and found that our coding mechanism on an average compresses 67.34% more than the other techniques.

## 1. INTRODUCTION

Information Retrieval (IR) refers to the processing of user requests, to obtain relevant information normally from unstructured textual data. The task of a full-text information retrieval system is to satisfy a user's information need by identifying the documents in a collection of documents that contain the desired information. This identification process requires a means of locating documents based on their content. A well known mechanism for providing such means is the inverted file index [1], [2]. An inverted file index consists of a record, or inverted list, for each term that appears in the document collection. A term's record contains an entry for every occurrence of the term in the document collection, identifying the document and possibly giving the location of the occurrence or a weight associated with the occurrence. Similarly an index may be a list of words and associated pointers to where those words can be found in a document. Indexes are designed to help the reader find information quickly and easily . Inverted file indices can become quite large. Some commercial systems contain millions of full text documents, occupying gigabytes of disk space. An inverted file index for such a collection will contain hundreds of thousands of records, ranging in size from just a few bytes to millions of bytes.

More over the B+ tree is a widely used index structure that maintains their efficiency irrespective of insertion and deletion of data [3]. Similar to B+-tree, B-tree allows search-key values to appear only once; eliminates redundant storage of search keys [3]. Another important indexing mechanism is hashing a hash function is computed on some attribute of each record; the result specifies in which block of the file the record should be placed. Bitmap indexes are special types of index designed for easy querying on multiple keys although each bitmap index is built on a single key [5].

Due to the enormous amount of data it is very much difficult to find the relevant information quickly. On the other hand as index is the main constituents of an IR system, it grows rapidly with the increase of data. Eventually compression of index structure has become the prime research issue. There are two more subtle benefits of compression. The first is increased use of caching. Search systems use some parts of the dictionary and the index much more than others [8]. The second more subtle advantage of compression is faster transfer of data from disk to memory. Efficient decompression algorithms run so fast on modern hardware that the total time of transferring a compressed chunk of data from disk and then decompressing it is usually less than transferring the same chunk of data in uncompressed form. So in most cases, the retrieval system will run faster on compressed postings lists than on uncompressed postings lists [9].

## 2. LITERATURE REVIEW

An efficient compression based index structure is developed by Justin Zobel, Alistair Moffat and Ron Sacks-Davis [1]. [1] Shows gamma technique which compresses 5.1% more efficiently then binary technique. They also show that $V_T$ compression technique compresses 4.5% more efficiently then binary. Another index structure is developed by Jinlin Chen, Terry Cook where they used d-gap method for index compression [2]. A comparison between d-gap and gamma is also shown in [2]. Maxim Martynov, Boris Novikov proposed an algorithm for query evaluation in text retrieval systems based on inverted lists, augmented with additional data structure and estimate expected performance gains [3]. This data structure is able to support dynamic indexing. A simple run-length compression method to use the codes for integers is described by Elias [4]. A research which shows indexing of very large data files containing hundreds of thousands or possibly millions of records is developed by [5]. Another compression based index structure is developed by Justin Zobel, Alistair Moffat and Ron Sacks-Davis [6]. A compressed inverted file index to search such a lexicon for entries that match a pattern or partially specified term is shown in [6]. This method provides an effective compression between speed and space.

## 3. COMPRESSING THE INDEX STRUCTURE

An inverted file index consists of a record, or inverted list, for each term that appears in the document collection. An inverted file index for such a collection will contain hundreds of thousands of records, ranging in size from just a few bytes to millions of bytes. So compression play important role in inverted file index.

### A. General structure of inverted file indexing

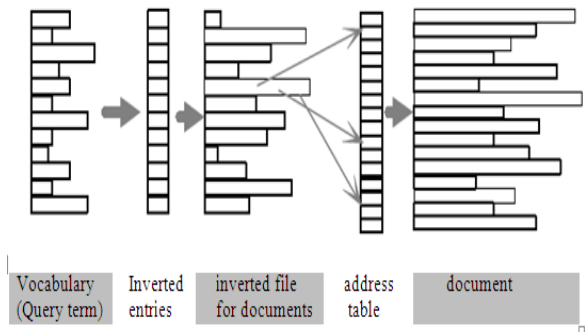

**Fig. 1. The general index structure**

A general inverted file index consists of two parts: a set of inverted file entries, being lists of identifiers of the records containing each indexed word; and a search structure or vocabulary for identifying the location of the inverted file entry for each word has shown in Fig. 1. We assume that inverted file entries store ordinal record numbers rather than addresses, and so to map the resulting record identifiers to disc addresses there must also be an address table (or disc mapping). Here documents are sequentially numbered and are accessible via addressing table. Inserting a new record means that each of the words in the record now occurs in one more record overall, thus changing the weight of each of these words. If record length was based on sums of weights, the length of every record containing any one of these words would have to be recomputed. In many applications it would be preferable to use a simpler measure of length, for which ranking performance may be worse but update is feasible.

### B. Compression on general index structure

For every word (Index Term) that occurs in a stored document, the inverted list contains an entry that includes a document number and optionally position information. For better search speed and storage utilization the inverted lists are ordered to enable run-length encoding. The general inverted file entries are shown in Table I.

**Table 1: General Inverted File Entries**

| Term | Document number | Weight |
|---|---|---|
| CSE | 20,58,222223,1111111 | 80,70,50,30 |
| ESRM | 90,50,21,5688,47584 | 85,40,30,20,15 |
| CPS | 50,199999,77777713 | 70,60,50 |
| BGE | 5555555,12 | 80,60 |
| FTNS | 2855555,233333 | 90,70 |
| computer | 124,5848,66687 | 95,40,30 |
| Book | 82,3333333,22222 | 80,60,20 |
| pen | 10000000,12,65,24 | 70,50,40,30 |

The inverted file size is of an uncompressed record-level inverted file, assuming a binary code of $\lceil \log_2 N \rceil$ bits per records identifier where N is the number of records in the collections. Rather than compressing the series of record numbers in an inverted file entry, we compression is on done their run-length encoding, the series of differences between successive numbers [3].
For example, the inverted file entry

50,70,110,190,240,…..
has the run-length encoding

50,20,40,80,50,………

### C. Our proposed compression method on index structure

In this paper we propose a variation of these schemes with improved performance especially for queries with large number of index terms. The model of data structure for this indexing scheme is represented on Fig. 2. Here we can modify the content of general inverted indexing structure in two parts. First we changed the inverted file entries that are shown in Table II. Secondly we changed the address table for documents. Here we divided it into two parts, the first part (1) contains the document number, which can not be supported by compression technique, shown in Table III.

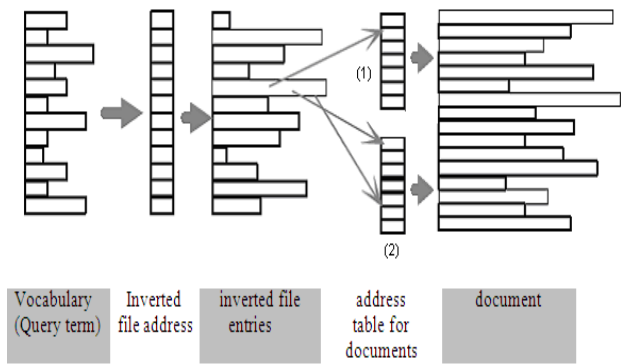

Fig. 2. Proposed inverted file index structure

The second part (2) contains the document number, which supports our compression technique, which is shown in Table IV. In inverted file entries our compression method works on the document number. We compressed only those numbers that supports our compressing algorithm, according to Table VI. For example original document number is 222222331 after using our compression algorithm it will be replaced by 2B331. Inverted file entries table is shown in above below. The converted document number of Table I will be like Table II using our algorithm. If a digit occurs more then 5 times sequentially in a document number then the occurrence number will be replaced by a code according to Table V. So, 222222331 will be 2B331. But during storing we will store the equivalent coded value of 101011001100110001 using Table VI instead of normal binary value 110100111110110101111111011. It causes a significant amount of compression for large document number. The proposed algorithm is given in Fig. 3. This compression technique gives the efficient facility of decoding also. During decoding we will scan the encode binary value of the compressed document number from right and find its equivalent value using Table VI.

| Term | Document number | Weight |
|---|---|---|
| CSE | 20,58,2A3,1C | 80,70,50,30 |
| ESRM | 90,50,21,5688,47584 | 85,40,30,20,15 |
| CPS | 50,19A,7B13 | 70,60,50 |
| BGE | 5C,12 | 80,60 |
| FTNS | 285A,23A | 90,70 |
| computer | 124,5848,66687 | 95,40,30 |
| Book | 82,3C,2A | 80,60,20 |
| pen | 10C,12,65,24 | 70,50,40,30 |

Table 2: Proposed Inverted File Entries

In our work we divided the address table for document into two parts. As a result the overall time for finding relevant information against a query has been reduced. The document number that does not support our algorithm scan the address table in Table III. Otherwise it search in Table IV. This mechanism reduced the searching time for finding relevant information.

Table 3 Address Table For Documents Part (1)

| Document no | Address | |
|---|---|---|
| 1 | | Doc#1 |
| 2 | | Doc#2 |
| 3 | | Doc#1111 |
| . | . | |
| . | . | |
| 11110 | | Doc#1111 |
| 11112 | | Doc#1111 |
| . | . | |

Table 4 Address Table For Documents Part (2)

| Document no | Address | |
|---|---|---|
| 1A | | Doc#1A |
| 2A | | Doc#2A |
| 3A | | Doc#3A |
| 4A | | Doc#4A |
| . | . | |
| . | . | |

Table V shows the corresponding coded value of original value having value greater then 3. Here we assigned the value up to 100.

Compression Algorithm:

1. Input a document number.
2. Scan a digit from that number.
3. If this digit similar to previous digit or first one.
4.     Increase counter by 1.
5. Else if counter is greater then or equal to 5.
6.     Add the digit to the variable and add the corresponding code of (counter – 1) From the Table V.
7. Else If counter is less then 5.
8.     Add the digit to the variable.
9. Output the variable as compressed code.

Fig. 3. Our proposed algorithm

**Table 5 Coding Of Frequency Count for Similar Digits**

| Original value | Coded value |
|---|---|
| 4 | A |
| 5 | B |
| 6 | C |
| 7 | D |
| 8 | E |
| 9 | F |
| 10 | AA |
| . | . |
| 17 | BC |
| . | . |
| . | . |
| 81 | AFF |
| . | . |
| . | . |
| 99 | BCE |
| 100 | BCF |

In our experiment we will store our coded value in the database according to Table 6

**Table 6 Assigning Code**

| Coded Value | Assigned Code |
|---|---|
| 0 | 0 |
| 1 | 1 |
| 2 | 10 |
| 3 | 11 |
| 4 | 100 |
| 5 | 101 |
| 6 | 110 |
| 7 | 111 |
| 8 | 1000 |
| 9 | 1001 |
| A | 1010 |
| B | 1011 |
| C | 1100 |
| D | 1101 |
| E | 1110 |
| F | 1111 |

## 4. EXPERIMENT AND RESULT

Here we experimented the corresponding bit representation of the original document number and our compressed document number. For our compression purpose we used the hexadecimal coding technique of Table VI. Our algorithm works good for large document numbers. For huge data set of different domain it is obvious that there may be large document number. We analyzed the IR system of our university library data set. For document numbers we experimented our algorithm. For large document number our algorithm shows significant compression ratio. Different compression results are shown in Table VII and Table VIII. Table VII shows that our propose algorithm perform 56.84% more compression than binary coding. We can find 77.85% more compression than gamma coding [4]. Fig. 4. Shows the graphical representation of our compression technique to other techniques.

**Table 7 Compression Using Binary Code**

| Document number | Binary representation | Bit representation of our compressed value | % of compression to binary |
|---|---|---|---|
| 55555 | 1101100100000011 (16 bit) | 1011010 (7 bit) | 56.25 |
| 999999 | 11110100001000111111 (20 bit) | 10011011 (8 bit) | 60 |
| 1322222 | 101000010110011101110 (21 bit) | 1001100101010 (13 bit) | 38.09 |
| 1888888 | 111001110001001111000 (21 bit) | 110001011 (9 bit) | 57.14 |
| 2222222 | 1000011111000010001110 (22 bit) | 101100 (6 bit) | 72.72 |

**Table 8 Compression Using Gamma Code**

| Document number | Gamma representation | Bit representation of our compressed value | % of compression to gamma |
|---|---|---|---|
| 55555 | 111111111111111101011001000000101 (31 bit) | 1011010 (7 bit) | 56.25 |
| 999999 | 11111111111111111110111010000100011111 (39 bit) | 10011011 (8 bit) | 60 |
| 1322222 | 1111111111111111111110010000101100111011110 (41 bit) | 1001100101010 (13 bit) | 38.09 |
| 1888888 | 11111111111111111111101100110100100111100 (41 bit) | 110001011 (9 bit) | 57.14 |
| 2222222 | 1111111111111111111110100001111011111000110 (43bit) | 101100 (6 bit) | 72.72 |

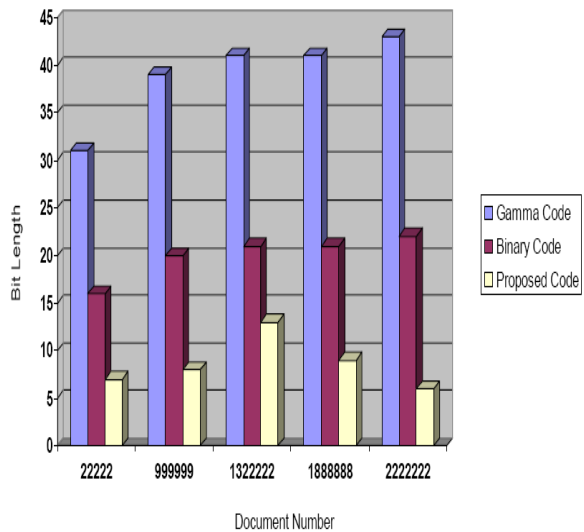

**Fig. 4. Comparison with Gamma code and Binary code**

## 5. CONCLUDING REMARKS

In this paper we used new algorithm for compressing inverted index. Although our proposed technique performs better results on huge number of documents only, it shows significant compression ratio to the well – known gamma and binary technique. Decompressing the encoded characters is also very straight forward in our algorithm. On the other hand as we split the index structure into two parts, it also gives a comparatively quick response to any user query.

## REFERENCES

[1] Justin Zobel, Alistair Moffat, and Ron Sacks-Davis, "Efficient indexing technique for full-text database systems", *In Proc. 18$^{th}$ Intnl.Conf. on VLDB*, pp. 352–362, 1992.

[2] Jinlin Chen, Terry Cook, "Using d-gap Patterns for Index Compression", *In Proc. of the WWW2007*.

[3] Maxim Martynov, Boris Novikov, "An Indexing Algorithm for Text Retrieval", *The International Workshop on Advances in Databases and Information Systems, Moscow, September 1996.*

[4] P. Elias, "Universal codeword sets and representation of the integers", *IEEE Trans. On Info. Theory*, pp. 194–203, February 1975.

[5] A.J. Kent, R. Sacks-Davis, and K. Ramamohanarao, "A signature file scheme based on multiple organizations for indexing very large text databases", *American Society for Inf. Science,* pp. 508-534, 1990.

[6] Justin Zobel, Alistair Moffat, and Ron Sacks-Davis, "Searching large lexicons for partially specified terms using compressed Inverted files", *In Proc.19th Intnl.Conf. on VLDB,* pp. 290–301, 1992.

[7] V. N. Anh and A. Moffat , "Index compression using fixed binary codeword's", *In Proc. 15th Australasian Database Conference,* pp. 61-67, 2004

[8] J. Zobel and A. Moffat, "Adding compression to a full-text retrieval system", *In Proc. 15'Th Australian Computer Science Conference,* pp. 1077-1089, January 1992.

[9] I. H. Witten, T.C. Bell, and C. Nevill, "Models for compression in full-text retrieval systems", *Proc. IEEE Data Compression Conference,* pp. 23-32, April 1991.